\documentclass[useAMS,usenatbib, use]{mn2e}
\usepackage{epsfig, color, amsmath}
\title{Bayesian Seismology of the Sun}
\author[Gruberbauer and Guenther]{M. Gruberbauer$^{1}$\thanks{E-mail: mgruberbauer@ap.smu.ca}, 
D.B. Guenther$^{1}$\thanks{E-mail: guenther@ap.smu.ca}
\\
$^{1}$Institute for Computational Astrophysics, Department of Astronomy and Physics, Saint Mary's University, Halifax, NS B3H 3C3, Canada\\}

\newcommand{\XO}{X_{\rm 0}}
\newcommand{\ZO}{Z_{\rm 0}}
\newcommand{\YO}{Y_{\rm 0}}

\newcommand{\Zs}{Z_{\rm s}}
\newcommand{\Ys}{Y_{\rm s}}
\newcommand{\ZoX}{Z_s/X_s}
\newcommand{\ZoXsun}{(Z_s/X_s)_{\odot}}
\newcommand{\aml}{\alpha_{\rm ml}}
\newcommand{\muHz}{\mu{\rm Hz}}

\begin{document}

\date{Accepted  Received}

\pagerange{\pageref{firstpage}--\pageref{lastpage}} \pubyear{2013}

\maketitle

\label{firstpage}

\begin{abstract}
We perform a Bayesian grid-based analysis of the solar {\it l}=0,1,2 and 3 p modes obtained via BiSON in order to deliver the first Bayesian asteroseismic analysis of the solar composition problem. We do not find decisive evidence to prefer either of the contending chemical compositions, although the revised solar abundances (AGSS09) are more probable in general. We do find indications for systematic problems in standard stellar evolution models, unrelated to the consequences of inadequate modelling of the outer layers on the higher-order modes. The seismic observables are best fit by solar models that are several hundred million years older than the meteoritic age of the Sun. Similarly, meteoritic age calibrated models do not adequately reproduce the observed seismic observables. Our results suggest that these problems will affect any asteroseismic inference that relies on a calibration to the Sun.

\end{abstract}


\section{Introduction}
\label{sec:intro}
The study of solar-type pulsation with its reliance on scaling relations \citep[e.g.,][]{huber2011} and calibrations of fundamental free parameters in stellar models (i.e., mixing length parameter and helium abundance) is ultimately anchored by what we know about the Sun and by how well seismology performs at identifying the Sun's key properties. Recent asteroseismic investigations of sun-like pulsators \citep[e.g.,][]{metcalfe2010, mathur2012} are able to give precise model-dependent constraints but it is difficult to assess their accuracy. Inferences from certain asteroseismic observables are not necessarily model dependent as can be verified using spectroscopy or interferometry \citep[e.g.,][]{huber2012}. However, full asteroseismic analyses that determine stellar ages and compositions, or decide among different implementations of how to model important physical processes (e.g., different approaches to convection) rely on a thorough calibration of the properties and parameters of the model. 

Several incompatibilities remain between solar modelling and the results inferred from helioseismology that can potentially affect our calibrations \citep[for a recent comprehensive review see][]{jcd2009}. For example, many investigators find that models that use the previous generation \citep{grevesse98} abundances fit helioseismic observables better than the current revised solar abundances \citep{asplund2005, asplund2009}. Consequently, the helium abundance and the resulting value for the ratio of the metal mass fraction to hydrogen mass fraction at the surface, $\ZoXsun$, is uncertain. We do know that inadequate modelling of the outer layers leads to the so-called ``surface effects" \citep[see, e.g.,][]{kjeldsen2008, gruberbauer2012} that worsens the model fit to higher order frequencies. Uncertainties in opacities, equations of state, nuclear reaction rates, and other global parameters also influence the properties of the solar model and, as a consequence, its seismic calibration. Recently, for instance, an increase in the opacities \citep{serenelli09} and different accretion scenarios \citep{serenelli11} have been identified as possible remedies for the disagreement between the results of helioseismic inversion and models based on the previous and current generation of chemical compositions.

Previous studies testing different model configurations, for example, different chemical mixtures, have often relied on the direct comparison of non-seismic observables and general properties inferred from helioseismology to stellar models calibrated to the non-seismic observables: age, radius, mass, luminosity, and in some cases also surface abundances. More recent approaches \citep{basu2007, chaplin2007, serenelli09} compared low-degree p modes, or rather various spacings derived from them, to models with solar characteristics. The result again suggests that they cannot be reconciled with the revised solar abundances. \cite{houdek2011} also developed an approach that uses quantities derived from the observed modes to infer solar model properties via iterative calibration procedures.

What is missing, though, is a test of the solar model with a tool that takes into account all the information given by the low-degree solar p modes and other constraints and which then results in a quantitative comparison of how much certain model properties are actually preferred on a global, probabilistic level. 
In our previous paper \citep[][hereafter Paper I]{gruberbauer2012} we introduced a new Bayesian method that uses prior information and properly treats known systematic effects (i.e., ``surface effects"). We performed a state-of-the-art, albeit, abbreviated grid-based asteroseismic analysis of the solar model.
In this paper we build on and extend our solar modelling by testing various chemical compositions and nuclear reaction rates.
Our goal is to answer the following questions:
\begin{enumerate}
\item Which models fit the solar modes and other observables the best?
\item Is there a clear preference for any of the chemical compositions and reaction rates?
\item How do the surface effects affect the fit?
\item How do our results affect the calibrations for asteroseismology?
\end{enumerate}

We approach our analysis, leaning more toward utilising the techniques applicable to asteroseismology than helioseismology. Specifically we will only utilise the lower {\it l}-valued p modes and we will allow all parameters except for the mass to remain unconstrained. We are, therefore, setting out to model the global properties of the Sun as a star, hence, to perform asteroseismology of the Sun.

\section{Grid-based fitting approach}

\subsection{Observations}
As in Paper I, we fit our models to the activity-corrected solar $l=0$, 1, 2, and 3 p modes obtained by using BiSON data \citep{broomhall2009}.
For our prior probabilities on other solar observables, we take an investigative approach by using both broad and narrow priors for the most important solar quantities: $T_{\rm eff}$, $L$, and age. This will help us to study the systematic dependencies of our results on the imposed constraints. For the general properties of the Sun, we use both a broad prior with $\log T_{\rm eff} = 3.7617 \pm 0.01$, and $\log\left(L/L_{\odot}\right) = 0.00 \pm 0.01$, or alternatively a more realistic but still conservative prior with $\log T_{\rm eff} = 3.7617 \pm 0.002$, and $\log\left(L/L_{\odot}\right) = 0.00 \pm 0.002$. Here $L_{\odot} = 3.8515 \pm 0.009 \cdot 10^{33} \rm erg\,s^{-1}$ (the average of the ERB-Nimbus and SMM/ARCRIM measurements; \cite{hickey1983}). For the solar mass, we use $M_{\odot} = 1.9891 \pm 0.0004 \cdot 10^{33}\rm g$ \citep{cohen1986}. As a reference for the solar age we take the result from \cite{bouvier2010} who determined a meteoritic age of the solar system of $\tau \approx 4.5682 \,\rm{Gyrs}$. As will be discussed in Section\,\ref{sec:ageprior}, we construct various uniform priors to allow a range of ages centred on this value. Finally, we also use the helioseismically inferred value of the radius of the base of the convection zone, $R_{\rm BCZ} = 0.713 \pm 0.001\,{R_{\odot}}$ \citep{jcd1991, basu97}. All priors are assumed to be normal distributions.

\subsection{Model physics}
\label{sec:model}

Just as in Paper I, our aim was to employ YREC \citep{demarque2008} and produce a set of dense grids covering a wide range in initial hydrogen mass fractions $\XO$, initial metal mass fractions $\ZO$, and mixing length parameters $\aml$. For this study we kept all model masses constrained to $1 \,M_{\odot}$, but we additionally varied the chemical composition and the nuclear reaction rates.

Our model tracks begin as completely convective Lane-Emden spheres \citep{lane1869, chandra1957} and are evolved from the Hayashi track \citep{hayashi1961} through the zero-age-main-sequence (ZAMS) to 6 Gyrs with each track consisting of approximately 2500 models. Only models between 4.0 and 6.0 Gyrs are included in the model grid. Constitutive physics include the OPAL98 \citep{iglesias1996} and \cite{alexander1994} opacity tables, as well as the Lawrence Livermore 2005 equation of state tables \citep{rogers1986, rogers1996}. Convective energy transport was modelled using the B\"ohm-Vitense mixing-length theory \citep{boehm1958}. The atmosphere model follows the ($T$-$\tau$) relation by \cite{krishna1966}. For each grid, we varied the chemical composition and tested two different nuclear reaction rates. We considered the GS98 mixture \citep{grevesse98}, the AGS05 mixture \citep{asplund2005}, and the AGSS09 mixture \citep{asplund2009}.
Nuclear reaction cross-sections were taken from \cite{bahcall2001} and the nuclear reaction rates from Table 21 in \cite{bahcall1988}. In addition, we also calculated grids using the NACRE rates \citep{angulo1999}. The effects of helium and heavy element diffusion \citep{bahcall1995} were included. Note that our atmosphere models and diffusion effects have been shown to require a larger value of mixing length parameter ($\aml \approx 2.0 - 2.2$) than standard Eddington atmospheres ($\aml \approx 1.7 - 1.8$) \citep{guenther1993}. The model grid spans: $\XO$ from 0.68 to 0.74 in steps of 0.01, $\ZO$ from 0.014 to 0.026 in steps of 0.001, and $\aml$ from 1.3 to 2.5 in steps of 0.1. 

The pulsation spectra were computed using the stellar pulsation code of \cite{guenther1994}, which solves the linearized, non-radial, non-adiabatic pulsation equations using the Henyey relaxation method. The non-adiabatic solutions include radiative energy gains and losses but do not include the effects of convection. We estimate the random $1\sigma$ uncertainties of our model frequencies to be of the order of $0.1\,\mu\rm Hz$. These uncertainties are properly propagated into all further calculations.

\subsection{Fitting method}
\label{sec:fitting}

Our Bayesian fitting method is explained in detail in Paper I. To briefly summarize, we compare theoretical and observed frequencies by calculating the likelihood that the two values agree were it not for the presence of random and systematic errors. These likelihoods are then combined using the sum rule and product rules of probability theory, and weighted by priors to arrive at correctly normalised probabilities. The random errors are assumed to be independent and Gaussian. Although frequency uncertainties are likely to be somewhat correlated depending on the data set quality and extraction technique, independence is a fundamental necessity to allow the independent treatment of surface effects. In the solar case the observational uncertainties are rather small, and so random errors in the model frequencies due to the model shell resolution ($\sim 0.1\,\muHz$), the influence of priors, and the surface effect treatment will outweigh the influence of correlations\footnote{Moreover, if frequency errors are derived from their marginal distribution as in Bayesian peak-bagging \citep[e.g.,][]{gruberbauer09, handberg11}, they can be treated as independent.}. The systematic errors in the case of solar-like stars are assumed to be similar to ``surface effects". At higher orders, observed frequencies are systematically lower than model frequencies, and the absolute frequency differences increase with frequency. This is modelled by introducing a systematic difference parameter, $\Delta$, between observed and calculated frequency so that

\begin{equation}
f_{\rm obs, i} = f_{\rm calc, i} + \gamma \Delta_i.
\end{equation}

In the case of surface effects, $\gamma = -1$. These $\Delta_i$ are then allowed to become larger at higher frequencies. The upper limit at each frequency is determined by the large frequency separation and a power law similar to the standard correction introduced by \cite{kjeldsen2008}. The $\Delta$ parameter is incorporated in a completely Bayesian fashion, using a $\beta$ prior to prefer smaller values over larger ones (see Paper I for more details). In addition, we always allow for the possibility that a mode is not significantly affected by any kind of systematic error. Altogether, this allows us to fully propagate uncertainties originating from the surface effects into all our results, and at the same time gives us more flexibility than the standard surface-effect correction.

We obtain probabilities for every evolutionary track in our grids, and within the tracks also for every model. We also obtain the correctly propagated distributions for systematic errors so that the model-dependent surface effect can be measured. In order to fully resolve the changes in stellar parameters and details in the stellar-model mode spectra, we oversample the evolutionary tracks via linear interpolation until the (normalised) probabilities no longer change significantly. Eventually, we obtain so-called evidence values, equivalent to the prior-weighted average likelihood, for every grid as a whole. These evidence values are also identical to the likelihood of the data (i.e., the solar frequency values) given the particular grids as conditional hypotheses. Just as the likelihoods for individual stellar models or, one step further, for evolutionary tracks can be used to compare their probabilities and evaluate the stellar parameters, the evidence values as likelihoods for whole grids can therefore be employed to perform a quantitative comparison between different input physics used in the grids\footnote{Other hypothesis modifications (e.g, different shapes of systematic errors) can in principle also be compared.}. This exemplifies the hierarchical structure of Bayesian analysis, which is discussed in more detail in Paper I and also in the more general literature \citep[e.g.,][]{gregory2005}.

\subsection{Analysis procedure}

The advantage of the Bayesian analysis method from Paper I is that many different approaches to fitting the same data set can be compared using the evidence values. Our goal is to see if there is a strong preference for either the GS98, AGS05, or AGSS09 mixture. We also want to test whether or not the NACRE nuclear reaction rates are an improvement. The corresponding grids will be designated as GS98N, AGS05N, and AGSS09N\footnote{Statements that are valid for both reaction rates will refer to both grids at once using the notation GS98(N), AGS05(N), or AGSS09(N)}. We use priors for the HRD position and age, as well as $R_{\rm BCZ}$, to see which of these grids are more consistent with well-known solar properties other than the frequencies. Also, by turning off the priors we can tell which solar-mass models best reproduce the frequencies irrespective of their fundamental parameters.

We will start our analysis without any priors and successively increase the prior information we use, to answer the questions outlined in Section\,\ref{sec:intro}. For example, if we were to find that the best solar models are much too old and luminous, or if the evidence values decrease when the priors are turned on, we will then have evidence that the model physics cannot reproduce an accurately calibrated solar model. 

It should be noted that all results presented in the following sections are highly dependent on the models used (i.e., what was described in Section\,\ref{sec:model}). We therefore cannot claim that our results represent the real Sun, as indeed we perform our analysis to investigate the similarities and systematic differences between models and observations. However, as explained in Paper I, our approach is capable to compare different grids produced from different codes and thus draw probabilistic inferences about systematic differences between these codes as well.

\section{Results}
\label{sec:results}

\subsection{No priors}

For the first test we did not use any of the luminosity, temperature, age, or $R_{\rm BCZ}$ constraints as formal priors. In this case the only effective prior is provided by the selection of the model grid parameters and the restriction to solar-mass models, hence, every model in the grids was given equal prior weight.

\begin{figure}
\centering
\includegraphics[width=\columnwidth]{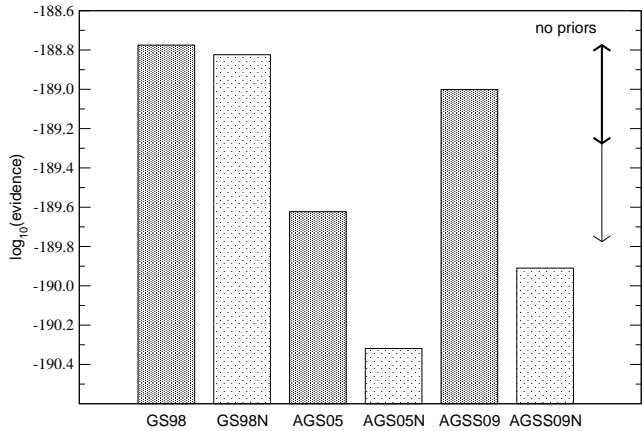}
\caption{Model grid performance without HRD, age or $R_{\rm BCZ}$ priors. The thick double-sided and thin arrows indicate strength of evidence that is ``barely worth mentioning" and ``substantial" respectively. Differences larger than the thin arrow can be considered ``strong" evidence (see text).}
\label{fig:nop_comp}
\end{figure}

Fig.\,\ref{fig:nop_comp} compares the grids in terms of the logarithm of the evidence. Note that differences between these values are equivalent to the logarithm of the posterior probability ratios for the grids as a whole under the condition that they have equal prior probabilities.

%(i.e., $P({\rm GS98} | I) = P({\rm AGS05} | I) = P({\rm AGSS09} | I) = ...$) . This means, for example,

%\begin{equation}
%\begin{split}
%\frac{\rm evidence_{GS98}}{\rm evidence_{AGS05}} = \frac{P(D | {\rm GS98}, I)}{P(D | {\rm AGS05}, I)} \\ = \frac{P({\rm GS98} | D, I)}{P({\rm AGS05} | D, I)} \frac{P({\rm AGS05} | I)}{P({\rm GS98} | I)}
%\end{split}
%\end{equation}

Following the guidelines provided by \cite{jeffreys61}, differences of up to 0.5 (or likelihood ratios up to 3) are considered ``barely worth mentioning". Differences between 0.5 and 1.0 indicate ``substantial" strength of evidence. Only when the differences rise above 1.0 (i.e., likelihood ratios $> 10$) should the strength of evidence be considered strong. Accordingly, the GS98-mixture models are not significantly better than AGSS09-mixture models. However, there is substantial evidence that the AGS05, AGS05N, and AGSS09N models do not reproduce the solar frequencies adequately, i.e., the GS98, GS98N, and AGSS09 are significantly better than the AGS05, AGS05N, and AGSS09N models. This indicates that there are problems with the AGS05 mixture and it also suggests that the NACRE rates have a detrimental effect on the model frequencies. Inspection of the frequencies for AGSS09N and AGSS05N reveals that, compared to the corresponding models in the AGSS09 and AGSS05 grids, the lower order modes do not fit as well and the surface effect also increases\footnote{As will be shown in Section\,\ref{sec:surf}, the former is usually more important than the latter.}. For instance, when  just considering the best evolutionary track for AGSS09, adopting the NACRE rates for the same track leads to decrease in probability by a factor of $\sim125$. The NACRE models are also older by $\sim 16$ Myrs and there is a significant increase in $R_{\rm BCZ}$ from 0.7164 to 0.7182.

\begin{figure}
\centering
\includegraphics[width=\columnwidth]{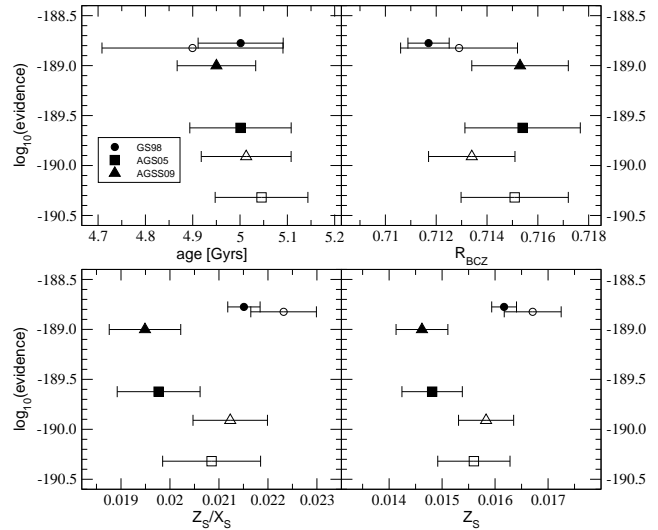}
\caption{Grid evidence versus mean values and uncertainties of some model properties when fitting the observed frequencies without any priors. Open symbols denote the corresponding NACRE grids.}
\label{fig:nop_par}
\end{figure}

In Fig.\,\ref{fig:nop_par} we show the mean values and uncertainties of some model properties, corresponding to the grids in Fig.\,\ref{fig:nop_comp}. Note that these uncertainties are caused by spreading the probabilities over a few different evolutionary tracks with models that  fit the frequencies best. If we were to restrict the parameter space by using priors as described in the next sections, then the probabilities will be mostly concentrated on only one or two evolutionary tracks and, consequently, the formal uncertainties will be reduced. 

Table\,\ref{tab:nop} contains more details for the most probable model parameters of the best and second-best evolutionary tracks in all grids. Considering the metallicities and the locations of the base of the convection zone, the results are similar to the general picture that has emerged in the literature. The GS98 and GS98N  models requires higher metallicities and a deeper base of the convection zone. Concerning the latter, the uncertainties are such that both AGSS09 and GS98(N) are in general agreement with $R_{\rm BCZ}$. None the less, the GS98(N) models fit this value a little bit better. Using the $R_{\rm BCZ}$ prior in the next sections will put a formal constraint on this as well.

It is disturbing, however, to see that all of the best models greatly overestimate the age of the Sun by several hundred million years. Furthermore, most of the models do not match the solar $T_{\rm eff}$ and luminosity very well. Therefore our next step is to ``switch on" either the broad or the more realistic priors constraining the Sun's position in the HR diagram.

\begin{table*}
   \centering
   \caption{Most probable parameters without priors. The quoted probabilities refer to the probability of the evolutionary track within each grid. $\XO$, $\ZO$: initial hydrogen and metal mass fractions; $\Zs$: metal mass fraction in the envelope; $R_{\rm BCZ}$: fractional radius of the base of the convection zone; $\alpha_{\rm ml}$: mixing length parameter.}
    \begin{tabular}{l c c c c c c c c c c c c} % Column formatting, @{} suppresses leading/trailing space
   \hline
   \hline
     grid & $T_{\rm eff} [K]$ & $L/L_{\odot}$ & $R/R_{\odot}$ & Age & $X_0$ & $Z_0$ & $Z_{\rm s}$ & $\ZoX$ & $R_{\rm BCZ}$ & $\alpha_{\rm ml}$  & Probability \\
     \hline
     GS98 & 5718 & 0.958 & 1.0001 & 5.022 & 0.72 & 0.018 & 0.0161 & 0.0214 & 0.7116 & 2.1 & 0.89 \\
               & 5802 & 1.016 & 0.9998 & 4.656 & 0.71 & 0.018 & 0.0162 & 0.0218 & 0.7139 & 2.2 & 0.05 \\
    \hline
     GS98N & 5660 & 0.920 & 1.0000 & 5.046 & 0.72 & 0.019 & 0.0170 & 0.0226 & 0.7114 & 2.0 & 0.52 \\
     	         & 5816 & 1.025 & 0.9997 & 4.637 & 0.71 & 0.018 & 0.0161 & 0.0217 & 0.7160 & 2.2 & 0.34 \\
     \hline        
     AGS05 & 5711 & 0.953 & 0.9997 & 4.967 & 0.72 & 0.016 & 0.0143 & 0.0190 & 0.7173 & 2.1 & 0.51 \\
     	          & 5754 & 0.983 & 1.0000 & 4.975 & 0.71 & 0.017 & 0.0152 & 0.0204 & 0.7139 & 2.2 & 0.38 \\
     \hline
     AGS05N & 5694 & 0.942 & 1.0000 & 5.041 & 0.71 & 0.018 & 0.0161 & 0.0216 & 0.7139 & 2.1 & 0.50 \\
     	         & 5647 & 0.911 & 0.9997 & 5.029 & 0.72 & 0.017 & 0.0152 & 0.0202 & 0.7165 & 2.0 & 0.26 \\
     \hline        
     AGSS09 & 5718 & 0.958 & 0.9998 & 4.932 & 0.72 & 0.016 & 0.0143 & 0.0190 & 0.7164 & 2.1 & 0.70 \\
       	           & 5761 & 0.988 & 1.0000 & 4.941 & 0.71 & 0.017 & 0.0152 & 0.0205 & 0.7132 & 2.2 & 0.26 \\
     \hline
     AGSS09N & 5701 & 0.947 & 1.0001 & 5.006 & 0.71 & 0.018 & 0.0161 & 0.0216 & 0.7128 & 2.1 & 0.67 \\
     	        & 5654 & 0.916 & 0.9998 & 4.993 & 0.72 & 0.017 & 0.0152 & 0.0202 & 0.7155 & 2.0 & 0.11 \\
     
    \hline     
   \end{tabular}
      \label{tab:nop}
\end{table*}

\subsection{$T_{\rm eff}$ and $L$ priors}

\begin{figure}
\centering
\includegraphics[width=\columnwidth]{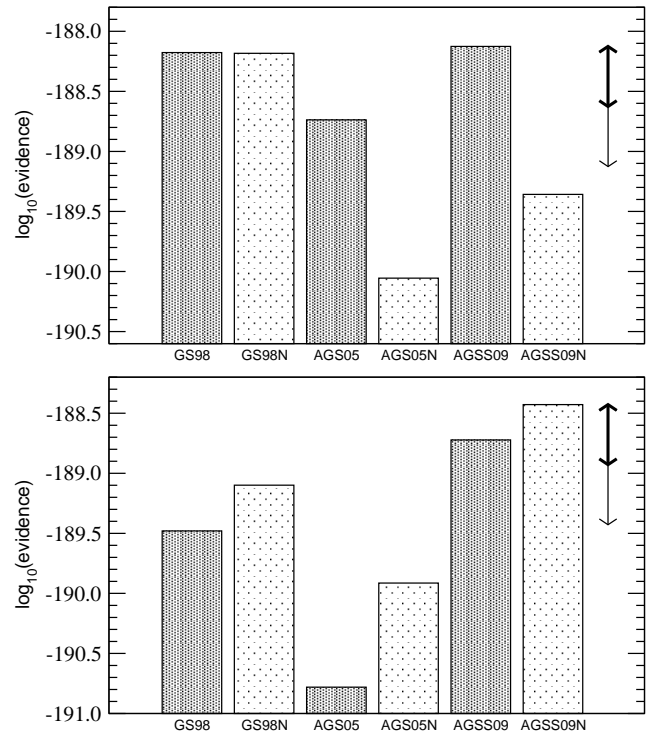}
\caption{Model grid performance with the broad (top panel) and the realistic (bottom panel) HRD prior.}
\label{fig:hrd_comp}
\end{figure}

As in Paper I we now use normal distributions as priors for $\log T_{\rm eff}$ and $\log\left(L/L_{\odot}\right)$ (hereafter: HRD prior). More weight is put on models that match the solar position in the HRD. Note that this does not mean that the best models will match the solar values. In this paper we employ slightly different HRD priors, using either a broad prior or a more realistic narrow prior based on current observational uncertainties. As we show below, the differences between the more realistic prior and the broad prior enable us to distinguish the chemical compositions. The resulting grid evidences are shown in the two panels of Fig.\,\ref{fig:hrd_comp}. 

For the broad HRD prior, an increase in evidence for all grids can be seen. This indicates that the models that are somewhat consistent with the solar values do include the majority of the best fit models. Since the evidence is a weighted average of the likelihood, however, most of the increase in evidence is caused by putting less weight on the many models that are clearly outside the solar values and do not match the solar frequencies at all. The relative likelihood ratios remain comparable to the ``no prior" case, but now AGSS09 is actually slightly more probable than the GS98(N) models. As before, the evidences of the three best grids are not different enough to clearly prefer one grid over the other. Table\,\ref{tab:hrd_broad} again gives information on the best fitting evolutionary tracks within each grid for the broad HRD prior. About half of the best or second-best models from the ``no prior" analysis remain among the most probable models but only GS98 shows the same models and ranking as before. It is interesting that the best-fitting model from the AGSS09 grid, which also is the overall best fit using the broad HRD prior, now matches the observed base of the convection zone closest from all models considered. Except for GS98N, the NACRE grids again perform worse than their counterparts. Note, however, that with the broad HRD prior the most probable basic model parameters are the same whether or not NACRE rates are used. 

For the realistic HRD prior, on the other hand, the GS98(N) grids receive an evidence penalty. Here, the preference for AGSS09(N) is more pronounced, and the previous decrease in evidence for AGSS09N is now compensated by its much closer match to the solar HRD position. As is shown in Table\,\ref{tab:hrd_real}, the most probable models for AGS05(N) and AGSS09(N) remain the same. However, the best AGS05 model underestimates luminosity and effective temperature and therefore its evidence decreases compared to the broad HRD prior.

\begin{table*}
   \centering
   \caption{Same as Table\,\ref{tab:nop} but with the broad HRD priors.}
    \begin{tabular}{l c c c c c c c c c c c c} % Column formatting, @{} suppresses leading/trailing space
   \hline
   \hline
     grid & $T_{\rm eff} [K]$ & $L/L_{\odot}$ & $R/R_{\odot}$ & Age & $X_0$ & $Z_0$ & $Z_{\rm s}$ & $\ZoX$ & $R_{\rm BCZ}$ & $\alpha_{\rm ml}$  & Probability \\
     \hline
     GS98 & 5718 & 0.958 & 1.0001 & 5.022 & 0.72 & 0.018 & 0.0161 & 0.0214 & 0.7116 & 2.1 & 0.77 \\
               & 5802 & 1.016 & 0.9998 & 4.656 & 0.71 & 0.018 & 0.0162 & 0.0218 & 0.7139 & 2.2 & 0.20 \\
    \hline
     GS98N & 5816 & 1.025 & 0.9997 & 4.637 & 0.71 & 0.018 & 0.0161 & 0.0217 & 0.7160 & 2.2 & 0.85 \\
     		 & 5732 & 0.967 & 1.0000 & 5.002 & 0.72 & 0.018 & 0.0161 & 0.0214 & 0.7127 & 2.1 & 0.10 \\
     \hline        
     AGS05 & 5754 & 0.983 & 1.0000 & 4.975 & 0.71 & 0.017 & 0.0152 & 0.0204 & 0.7139 & 2.2 & 0.84 \\
     		& 5711 & 0.953 & 0.9997 & 4.967 & 0.72 & 0.016 & 0.0143 & 0.0190 & 0.7173 & 2.1 & 0.15 \\
     \hline
     AGS05N & 5768 & 0.992 & 0.9999 & 4.957 & 0.71 & 0.017 & 0.0152 & 0.0204 & 0.7161 & 2.2 & 0.46 \\
     	         & 5725 & 0.962 & 0.9996 & 4.951 & 0.72 & 0.016 & 0.0143 & 0.0189 & 0.7189 & 2.1 & 0.36 \\
     \hline        
     AGSS09 & 5761 & 0.988 & 1.0000 & 4.941 & 0.71 & 0.017 & 0.0152 & 0.0205 & 0.7132 & 2.2 & 0.66 \\
     		   & 5718 & 0.958 & 0.9998 & 4.932 & 0.72 & 0.016 & 0.0143 & 0.0190 & 0.7164 & 2.1& 0.33 \\      	           
     \hline
     AGSS09N & 5775 & 0.997 & 1.0000 & 4.923 & 0.71 & 0.017 & 0.0152 & 0.0204 & 0.7149 & 2.2 & 0.69 \\
     	         & 5701 & 0.947 & 1.0001 & 5.006 & 0.71 & 0.018 & 0.0161 & 0.0216 & 0.7128 & 2.1 & 0.23 \\
     
    \hline     
   \end{tabular}
      \label{tab:hrd_broad}
\end{table*}

\begin{table*}
   \centering
   \caption{Same as Table\,\ref{tab:nop} but with the realistic HRD priors.}
    \begin{tabular}{l c c c c c c c c c c c c} % Column formatting, @{} suppresses leading/trailing space
   \hline
   \hline
     grid & $T_{\rm eff} [K]$ & $L/L_{\odot}$ & $R/R_{\odot}$ & Age & $X_0$ & $Z_0$ & $Z_{\rm s}$ & $\ZoX$ & $R_{\rm BCZ}$ & $\alpha_{\rm ml}$  & Probability \\
     \hline
     
     GS98 & 5767 & 0.992 & 1.0002 & 4.980 & 0.71 & 0.019 & 0.0170 & 0.0229 & 0.7096 & 2.2 & 0.83 \\
 	       & 5802 & 1.016 & 0.9998 & 4.656 & 0.71 & 0.018 & 0.0162 & 0.0218 & 0.7139 & 2.2 & 0.15 \\               
    \hline
     GS98N & 5780 & 1.001 & 1.0002 & 4.959 & 0.71 & 0.019 & 0.0170 & 0.0228 & 0.7109 & 2.2 & 0.997 \\
     		 & 5769 & 0.992 & 0.9995 & 4.660 & 0.72 & 0.017 & 0.0152 & 0.0203 & 0.7184 & 2.1 & 2.6e-3 \\
     \hline        
     AGS05 & 5754 & 0.983 & 1.0000 & 4.975 & 0.71 & 0.017 & 0.0152 & 0.0204 & 0.7139 & 2.2 & 0.90 \\
     		 & 5789 & 1.006 & 0.9995 & 4.848 & 0.72 & 0.015 & 0.0134 & 0.0178 & 0.7205 & 2.2 & 0.07 \\
     \hline
     AGS05N & 5768 & 0.992 & 0.9999 & 4.957 & 0.71 & 0.017 & 0.0152 & 0.0204 & 0.7161 & 2.2 & 0.99996 \\
     	         & 5779 & 1.000 & 0.9997 & 4.680 & 0.70 & 0.018 & 0.0161 & 0.0220 & 0.7177 & 2.2 & 3.7e-5 \\
     \hline        
     AGSS09 & 5761 & 0.988 & 1.0000 & 4.941 & 0.71 & 0.017 & 0.0152 & 0.0205 & 0.7132 & 2.2 & 0.9996 \\
     		   & 5796 & 1.011 & 0.9996 & 4.814 & 0.72 & 0.015 & 0.0134 & 0.0178 & 0.7191 & 2.2 & 3.0e-4 \\      	           
     \hline
     AGSS09N & 5775 & 0.997 & 1.0000 & 4.923 & 0.71 & 0.017 & 0.0152 & 0.0204 & 0.7149 & 2.2 & 0.999998 \\
     	       	  & 5787 & 1.005 & 0.9998 & 4.646 & 0.70 & 0.018 & 0.0161 & 0.0220 & 0.7158 & 2.2 & 1.0e-6 \\
     
    \hline     
   \end{tabular}
      \label{tab:hrd_real}
\end{table*}

All the conclusions drawn from the ``no prior" approach still apply, i.e., the model fits give us no clear indication for, e.g., preferring GS98(N) over AGSS09, but they do show significant evidence against AGS05 and for the detrimental effect of the NACRE rates.

\begin{figure}
\centering
\includegraphics[width=\columnwidth]{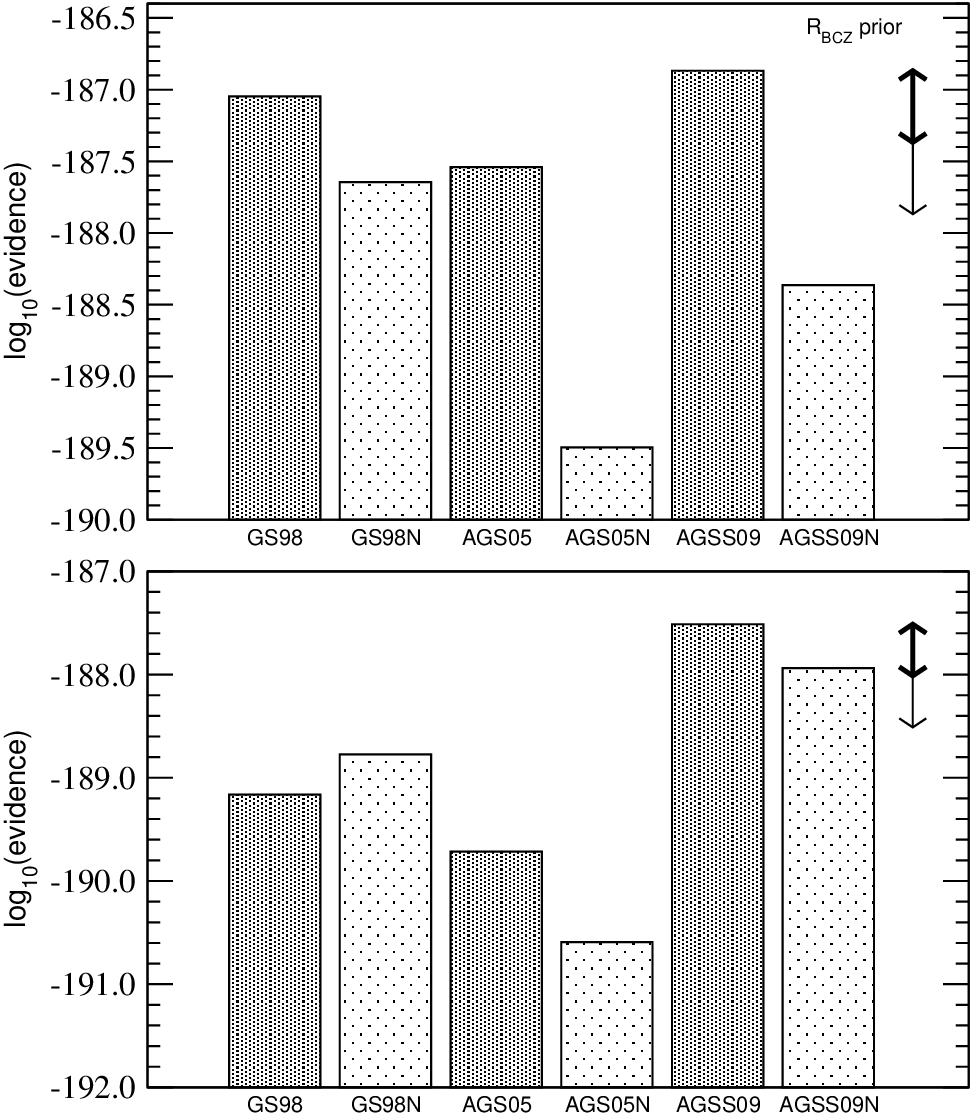}
\caption{Model grid performance with the $R_{\rm BCZ}$ prior, as well as the the broad (top panel) and the realistic (bottom panel) HRD prior.}
\label{fig:hrd_rbc_comp}
\end{figure}

Lastly, we turn on the $R_{\rm BCZ}$ prior in tandem with the HRD priors, which puts stronger constraints on a proper fit to the interior. The results are shown in Fig.\,\ref{fig:hrd_rbc_comp} and the corresponding model parameters for the realistic HRD prior are summarized in Table\,\ref{tab:hrd_real_rbc}.
Interestingly, for both HRD priors, AGSS09 manages to increase the probability contrast to the other models. The evidence rises once more, which signifies that the models that fit the pulsation frequencies also are among those that fit best to $R_{\rm BCZ}$. This is also confirmed by Table\,\ref{tab:hrd_real_rbc} which shows that the most probable models for AGSS09(N) and AGS05(N) have not changed. {\it For these mixtures the models that are best at reproducing the pulsation and broad HRD constraints also fit the realistic HRD constraints and the base of the convection zone.} This is also responsible for producing the enormous concentration of probability on the best evolutionary tracks. The bottom panel in Fig.\,\ref{fig:hrd_rbc_comp} also indicates that, with the realistic HRD prior and the $R_{\rm BCZ}$ constraint, there is formally strong evidence for the AGSS09 mixture to provide the overall most realistic solar model.

\begin{table*}
   \centering
   \caption{Same as Table\,\ref{tab:nop} but with the $R_{\rm BCZ}$ and realistic HRD priors.}
    \begin{tabular}{l c c c c c c c c c c c c} % Column formatting, @{} suppresses leading/trailing space
   \hline
   \hline
     grid & $T_{\rm eff} [K]$ & $L/L_{\odot}$ & $R/R_{\odot}$ & Age & $X_0$ & $Z_0$ & $Z_{\rm s}$ & $\ZoX$ & $R_{\rm BCZ}$ & $\alpha_{\rm ml}$  & Probability \\
     \hline
     
     GS98 & 5802 & 1.016 & 0.9998 & 4.656 & 0.71 & 0.018 & 0.0162 & 0.0218 & 0.7139 & 2.2 & 0.83 \\
 	       & 5789 & 1.007 & 1.0000 & 4.941 & 0.72 & 0.017 & 0.0152 & 0.0202 & 0.7130 & 2.2 & 0.14 \\               
    \hline
     GS98N & 5780 & 1.001 & 1.0002 & 4.959 & 0.71 & 0.019 & 0.0170 & 0.0228 & 0.7109 & 2.2 & 0.99998 \\
     		 & 5746 & 0.977 & 0.9998 & 4.694 & 0.71 & 0.019 & 0.0171 & 0.0230 & 0.7142 & 2.1 & 6.2e-6 \\
     \hline        
     AGS05 & 5754 & 0.983 & 1.0000 & 4.975 & 0.71 & 0.017 & 0.0152 & 0.0204 & 0.7139 & 2.2 & 0.99996 \\
     		 & 5798 & 1.014 & 1.0001 & 4.947 & 0.70 & 0.018 & 0.0161 & 0.0219 & 0.7119 & 2.3 & 1.8e-5 \\
     \hline
     AGS05N & 5768 & 0.992 & 0.9999 & 4.957 & 0.71 & 0.017 & 0.0152 & 0.0204 & 0.7161 & 2.2 & 0.9999997 \\
     	         & 5779 & 1.000 & 0.9997 & 4.680 & 0.70 & 0.018 & 0.0161 & 0.0220 & 0.7177 & 2.2 & 1.2e-7 \\
     \hline        
     AGSS09 & 5761 & 0.988 & 1.0000 & 4.941 & 0.71 & 0.017 & 0.0152 & 0.0205 & 0.7132 & 2.2 & 0.9999997 \\
     		   & 5773 & 0.996 & 0.9998 & 4.664 & 0.70 & 0.018 & 0.0162 & 0.0221 & 0.7139 & 2.2 & 2.6e-7 \\      	           
     \hline
     AGSS09N & 5775 & 0.997 & 1.0000 & 4.923 & 0.71 & 0.017 & 0.0152 & 0.0204 & 0.7149 & 2.2 & 0.9999998 \\
     	       	  & 5787 & 1.005 & 0.9998 & 4.646 & 0.70 & 0.018 & 0.0161 & 0.0220 & 0.7158 & 2.2 & 1.6e-7 \\
     
    \hline     
   \end{tabular}
      \label{tab:hrd_real_rbc}
\end{table*}

Nonetheless, all ages are still too high compared to the well-established meteoritic age estimate. We cannot consider these models to be properly calibrated to the Sun, even though the frequencies clearly prefer these solutions. We therefore now turn to age priors to avoid the solutions that are clearly too old (or too young).

\subsection{HRD and age priors}
\label{sec:ageprior}

In Paper I, we used a similar approach to rule out older models and employed a Gaussian prior centred on the meteoritic solar age but allowed for a few tens of millions of years of PMS evolution. In this paper, however, we chose to take a more careful approach. 

\begin{figure}
\includegraphics[width=0.9\columnwidth]{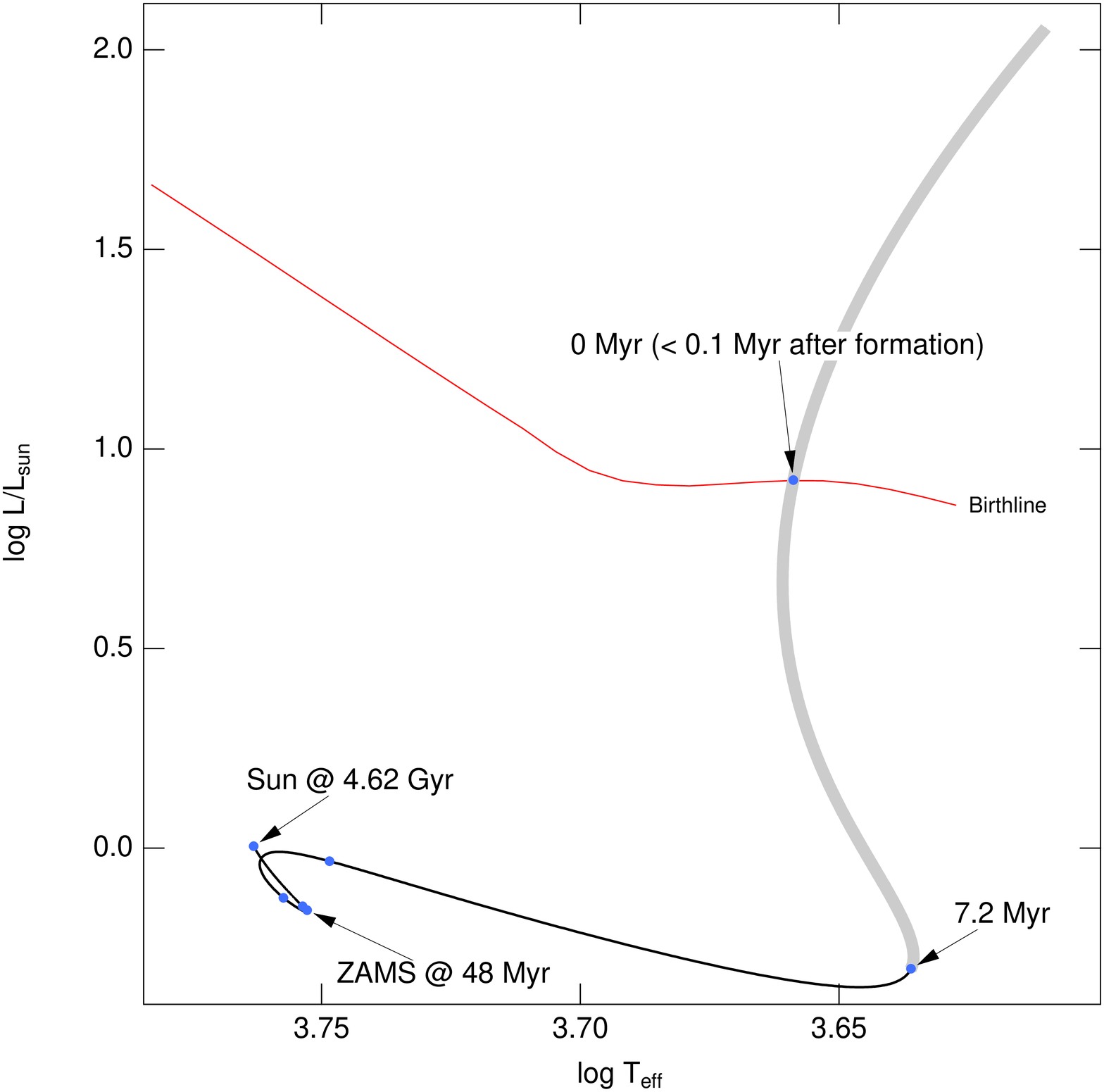}
\includegraphics[width=0.9\columnwidth]{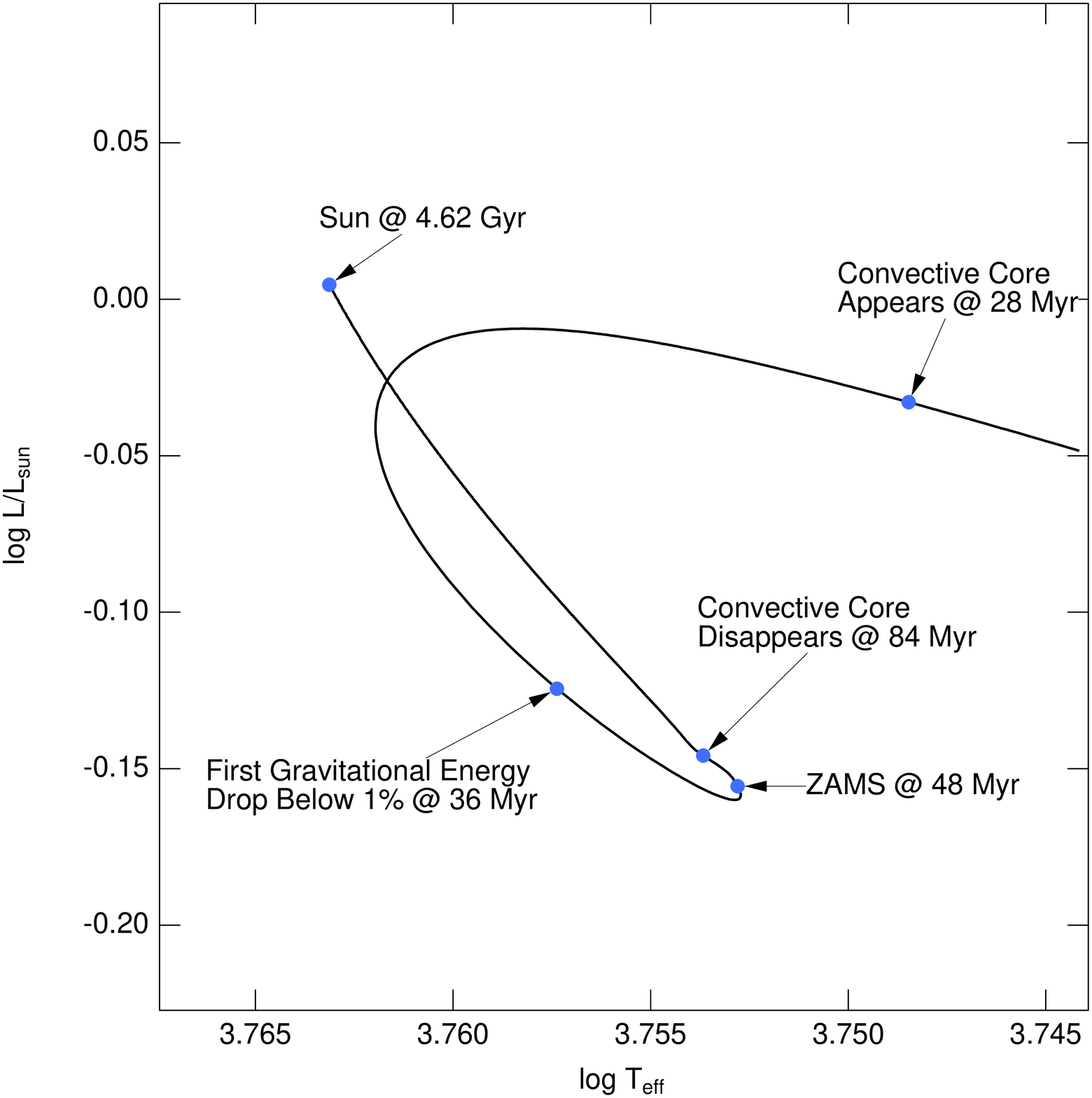}
\caption{Evolution of a one solar mass model. Evolution is started above the birthline. The model crosses the birthline after $< 0.1\,\rm Myrs$, at which point, the age of the model is reset to zero. The thick grey line indicates the period in which the primordial meteoritic material will cool and lock in the initial isotopic abundances used to date meteorites.}
\label{fig:pmsevo}
\end{figure}

Different authors often use different definitions for the age of their solar model (e.g., age from the birthline or age from the ZAMS). Therefore, Fig.\,\ref{fig:pmsevo} presents the age-related details in our solar model evolution. The meteoritic age is measured from the time when the initial abundance of the isotopes used to date the meteorites are no longer kept in equilibrium. This probably occurs at some point on the Hayashi track. We take the zero age of our models to coincide with the birthline as defined in \cite{palla1999}. This introduces an uncertainty of $\sim7 \,\rm Myrs$ between the meteoritic age and the birthline age, which is still smaller than the systematic errors in our model ages, which we estimate are of the order of a few tens of Myrs. Note, for example, that switching to the NACRE rates leads to a change in age of about $20 \,\rm Myrs$. 

In order to avoid putting too much weight on slight differences in the age, and to allow for systematic errors in the meteoritic age determination of perhaps a few Myrs, we will only use uniform age priors centred on the meteoritic age. The purpose of the age prior is therefore only to provide a cut-off for model ages above or below certain limits. We chose two different age priors, one more restrictive than the other, and we continue to use the HRD and $R_{\rm BCZ}$ priors.

\subsubsection{Broad age prior}

\begin{figure}
\centering
\includegraphics[width=\columnwidth]{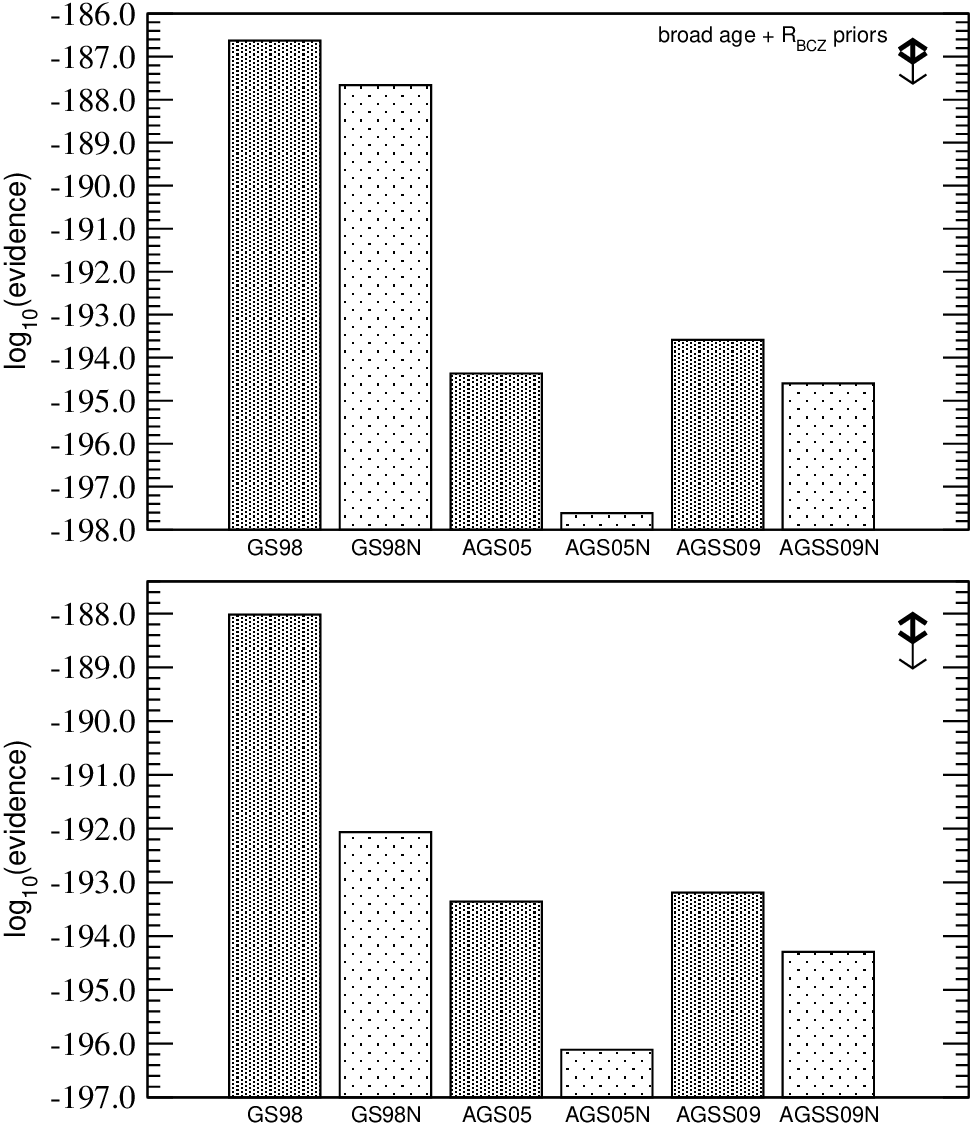}
\caption{Model grid performance with the $R_{\rm BCZ}$ and broad age prior, as well as the broad (top panel) and realistic (bottom panel) HRD prior.}
\label{fig:broad_rbc_comp}
\end{figure}

The broad age prior is a uniform prior that rules out very old or young models. We designed it to allow for an age range of 4.4 -- 4.7 Gyrs. This removes most of our previous best fits, but retains the good GS98(N) models which have $\approx 4.65$ Gyrs. Fig.\,\ref{fig:broad_rbc_comp} shows the results in terms of evidence. Clearly, the AGS05 and AGSS09 mixture have suffered a severe penalty for their older models are now outside the range allowed by the prior. The GS98 and GS98N models, on the other hand, show an increase in evidence compared to Fig.\,\ref{fig:hrd_comp} and therefore the evidence contrast has increased markedly. The realistic HRD prior does affect and slightly decrease this contrast, but since the AGS05(N) and AGSS09(N) grids have lost their previous best models to the age prior, the effect is not as pronounced as in Fig.\,\ref{fig:hrd_comp}. In terms of the strength of evidence, this result would amount to decisive evidence for the GS98(N) grids. Since the best models are the same for the broad and the realistic HRD prior, we only list the results for the latter in Table\,\ref{tab:bap_real_rbc}.

For GS98, the probability is now concentrated in the best model from Table\,\ref{tab:hrd_real_rbc}. Note that both the best GS98 and second best GS98N models have the same fundamental parameters, differing in their nuclear reaction rates. The AGS05(N) and AGSS09(N) grids all find the same basic model (except for the different mixture and reaction rates) with intermediate $\ZO = 0.018$ proving to be the most probable. 

Without the $R_{\rm BCZ}$ prior (not shown) the best GS98 and GS98N models are the same, and the overall evidence distribution is very similar as in Fig.\,\ref{fig:broad_rbc_comp}. However, the AGS05(N) and AGSS09(N) grids would prefer models with low metallicity ($\ZO = 0.016$) which produce values of $R_{\rm BCZ}$ well outside the range supported by the inversion results\footnote{These models will nonetheless turn out to be the most probable when we make the age constraint even stronger in the next section.}. For all grids, the ages of the best models are still too high by up to 150 Myrs.

%\begin{figure}
%\centering
%\includegraphics[width=\columnwidth]{figures/fig6}
%\caption{Model grid performance with the broad age prior, as well as the broad (top panel) and realistic (bottom panel) HRD prior.}
%\label{fig:broad_comp}
%\end{figure}

\begin{table*}
   \centering
   \caption{Same as Table\,\ref{tab:nop} but with $R_{\rm BCZ}$, realistic HRD and broad age priors.}
    \begin{tabular}{l c c c c c c c c c c c c} % Column formatting, @{} suppresses leading/trailing space
   \hline
   \hline
     grid & $T_{\rm eff} [K]$ & $L/L_{\odot}$ & $R/R_{\odot}$ & Age & $X_0$ & $Z_0$ & $Z_{\rm s}$ & $\ZoX$ & $R_{\rm BCZ}$ & $\alpha_{\rm ml}$  & Probability \\
     \hline
     GS98 & 5802 & 1.016 & 0.9998 & 4.656 & 0.71 & 0.018 & 0.0162 & 0.0218 & 0.7139 & 2.2 & 0.9998 \\
              & 5755 & 0.983 & 0.9996 & 4.678 & 0.72 & 0.017 & 0.0153 & 0.0203 & 0.7167 & 2.1 & 1.0e-4 \\
    \hline
     GS98N & 5746 & 0.977 & 0.9998 & 4.694 & 0.71 & 0.019 & 0.0171 & 0.0230 & 0.7142 & 2.1 & 0.43 \\
     		 & 5816 & 1.025 & 0.9997 & 4.637 & 0.71 & 0.018 & 0.0161 & 0.0217 & 0.7160 & 2.2 & 0.30 \\
     \hline        
     AGS05 & 5766 & 0.990 & 0.9997 & 4.697 & 0.70 & 0.018 & 0.0162 & 0.0220 & 0.7149 & 2.2 & 0.9999997 \\
     		& 5795 & 1.010 & 0.9994 & 4.613 & 0.71 & 0.016 & 0.0143 & 0.0193 & 0.7204 & 2.2 & 2.9e-7 \\
     \hline
     AGS05N & 5779 & 1.000 & 0.9997 & 4.680 & 0.70 & 0.018 & 0.0161 & 0.0220 & 0.7177 & 2.2 & 1.00 \\
     	         & 5733 & 0.968 & 0.9995 & 4.687 & 0.71 & 0.017 & 0.0152 & 0.0205 & 0.7194 & 2.1 & 1.8e-13 \\
     \hline        
     AGSS09 & 5773 & 0.996 & 0.9998 & 4.664 & 0.70 & 0.018 & 0.0162 & 0.0221 & 0.7139 & 2.2 & 0.999999 \\
     		   & 5802 & 1.015 & 0.9995 & 4.580 & 0.71 & 0.016 & 0.0144 & 0.0193 & 0.7192 & 2.2 & 9.9e-7 \\      	           
     \hline
     AGSS09N & 5787 & 1.005 & 0.9998 & 4.646 & 0.70 & 0.018 & 0.0161 & 0.0220 & 0.7158 & 2.2 & 0.999998 \\
     	         & 5767 & 0.991 & 0.9999 & 4.671 & 0.69 & 0.020 & 0.0180 & 0.0248 & 0.7121 & 2.2 & 1.0e-6 \\
     
    \hline     
   \end{tabular}
      \label{tab:bap_real_rbc}
\end{table*}

\subsubsection{Narrow age prior}

\begin{figure}
\centering
\includegraphics[width=\columnwidth]{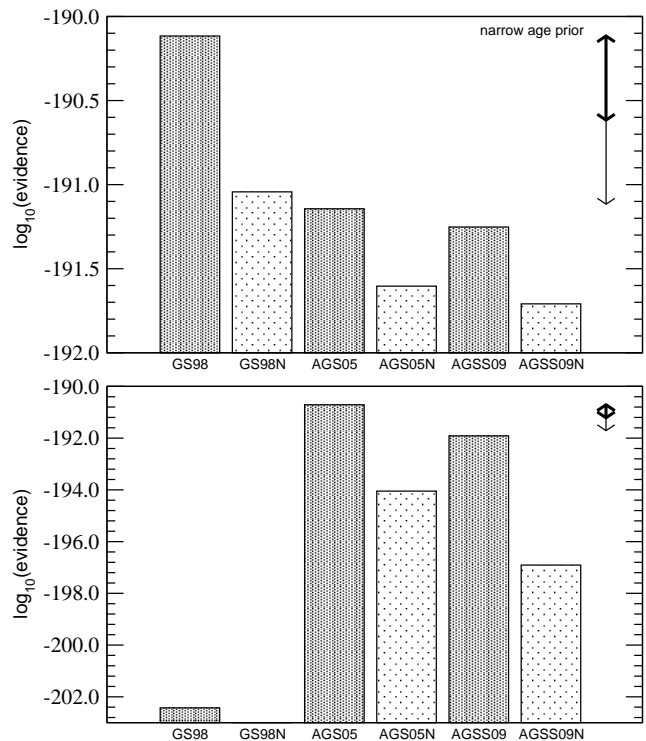}
\caption{Model grid performance with the narrow age prior, as well as the broad (top panel) and realistic (bottom panel) HRD prior.}
\label{fig:med_comp}
\end{figure}

In order to see how fully age-constrained solar models in the GS98(N) grids compare to the AGS04 and AGSS09 models, we restricted the age even further by employing a narrow uniform age prior that only allows ages of 4.52 -- 4.62 Gyrs. As shown in Fig.\,\ref{fig:med_comp}, the narrow age prior has a big effect on the analysis. Since it is interesting to see whether models at the correct age can fit the base of the convection zone, we first perform the analysis without the $R_{\rm BCZ}$ prior. 

Compared to Fig.\,\ref{fig:broad_rbc_comp} and for the broad HRD prior, the narrow age constraint strongly decreases the evidence for the GS98(N) models, while increasing the evidence for the other models. GS98 still comes out to be the most probable by an order of magnitude. The remaining grids show more or less comparable evidences, but AGS05N and AGSS09N are worse than their non-NACRE counterparts. All solutions for AGS05 and AGSS09 favour the same basic model parameters. Table\,\ref{tab:map_broad} lists the corresponding most probable models. Ultimately, the narrow age prior has led to models which are very close to the meteoritic solar age without constraining them too strongly (as would be the case for a non-uniform, e.g., Gaussian, age prior) so that we do not rule out completely the possibility of systematic errors in the stellar model age. All of the best models, irrespective of mixture or reaction rates, have $\XO = 0.71$, $\ZO < 0.018$, and $R_{\rm BCZ} > 0.716$. Compared to the revised mixtures, GS98(N) has slightly higher metallicities and requires a larger mixing length parameter. 

For the realistic HRD prior, however, the situation is completely different. All GS98(N) models drop out of the discussion due to a big decrease in evidence. The effective temperature and luminosity values (see, e.g., the results for GS98N in Table\,\ref{tab:map_broad}) are so far outside the prior range that the prior probability terms become close to our numerical threshold values. The only GS98(N) models that still retain what is left of the evidence have $\ZO < 0.017$. Even though the evidence picture has changed drastically, the actual best models for the AGS05 and AGSS09 grids are still the same as with the broad HRD prior.

\begin{figure}
\centering
\includegraphics[width=\columnwidth]{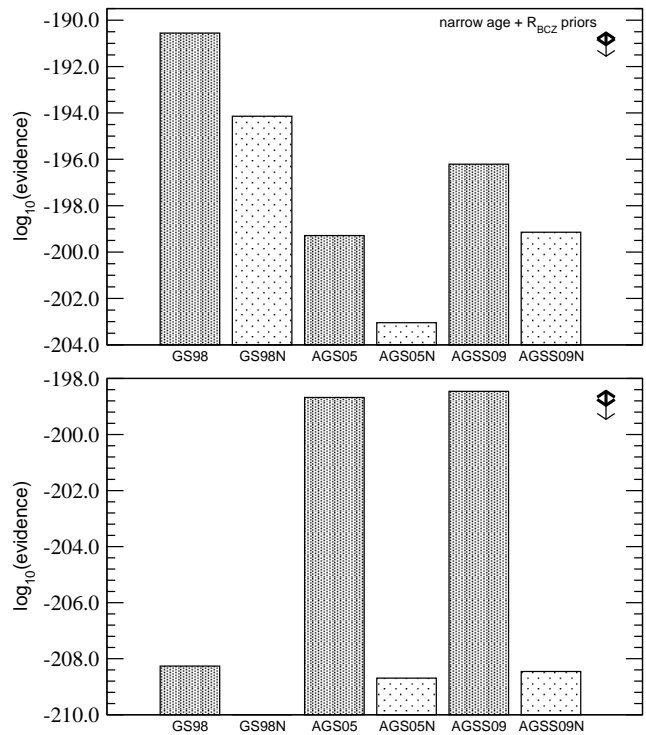}
\caption{Model grid performance with the $R_{\rm BCZ}$ and narrow age priors, as well as the broad (top panel) and realistic (bottom panel) HRD prior.}
\label{fig:med_rbc_comp}
\end{figure}

Finally, we again turn on the $R_{\rm BCZ}$ prior. The evidence results are depicted in Fig.\,\ref{fig:med_rbc_comp}. For the broad HRD prior, the evidence present a similar picture as before, but the contrast between GS98(N) and the AGS05(N) and AGSS09(N) models has intensified. Furthermore, the most probable models for GS98N, AGS05(N) and AGSS09(N) now have higher metal mass fraction as before. The overall most probable model of the GS98 grid, which far outweighs the others in terms of evidence, still is the same as in Table\,\ref{tab:map_broad}.

For the realistic HRD prior, which most closely reflects all of our prior knowledge of the Sun, the verdict is clear as well. However, for this prior, it is the AGS05 and the AGSS09 models which are preferred. {\it The evidence contrast between these two grids and the others is the highest contrast measured in all analyses performed in this paper}. The influence of the realistic HRD prior is again substantial and even produces a null result for the GS98N grid because of our numerical thresholds. The parameters for the most probable models are given in Table\,\ref{tab:map_real_rbc}. All best models now have $\XO < 0.73$, $\ZO = 0.016$, $\alpha_{\rm ml} = 2.2$, and $R_{\rm BCZ} > 0.719$, and the models with revised composition
agree on $\XO = 0.71$ as well.

\begin{table*}
   \centering
   \caption{Same as Table\,\ref{tab:nop} but with broad HRD and narrow age priors.}
    \begin{tabular}{l c c c c c c c c c c c c} % Column formatting, @{} suppresses leading/trailing space
   \hline
   \hline
     grid & $T_{\rm eff} [K]$ & $L/L_{\odot}$ & $R/R_{\odot}$ & Age & $X_0$ & $Z_0$ & $Z_{\rm s}$ & $\ZoX$ & $R_{\rm BCZ}$ & $\alpha_{\rm ml}$  & Probability \\
     \hline
     GS98 & 5872 & 1.065 & 0.9997 & 4.566 & 0.71 & 0.017 & 0.0153 & 0.0205 & 0.7162 & 2.3 & 0.98 \\
               & 5829 & 1.034 & 0.9995 & 4.591 & 0.72 & 0.016 & 0.0144 & 0.0191 & 0.7192 & 2.2 & 0.02 \\
    \hline
     GS98N & 5885 & 1.075 & 0.9996 & 4.547 & 0.71 & 0.017 & 0.0152 & 0.0205 & 0.7183 & 2.3 & 0.999 \\
     		 & 5843 & 1.043 & 0.9994 & 4.573 & 0.72 & 0.016 & 0.0143 & 0.0190 & 0.7212 & 2.2 & 4.3e-4 \\
     \hline        
     AGS05 & 5795 & 1.010 & 0.9994 & 4.613 & 0.71 & 0.016 & 0.0143 & 0.0193 & 0.7204 & 2.2 & 0.9999 \\
     		& 5837 & 1.040 & 0.9997 & 4.605 & 0.70 & 0.017 & 0.0153 & 0.0208 & 0.7180 & 2.3 & 1.6e-5 \\
     \hline
     AGS05N & 5809 & 1.019 & 0.9994 & 4.596 & 0.71 & 0.016 & 0.0143 & 0.0192 & 0.7220 & 2.2 & 0.9999 \\
     	         & 5851 & 1.050 & 0.9996 & 4.588 & 0.70 & 0.017 & 0.0152 & 0.0207 & 0.7194 & 2.3 & 2.5e-5 \\
     \hline        
     AGSS09 & 5802 & 1.015 & 0.9995 & 4.580 & 0.71 & 0.016 & 0.0144 & 0.0193 & 0.7192 & 2.2 & 0.9999 \\
     		   & 5845 & 1.045 & 0.9997 & 4.572 & 0.70 & 0.017 & 0.0153 & 0.0208 & 0.7162 & 2.3 & 2.5e-5 \\      	           
     \hline
     AGSS09N & 5816 & 1.024 & 0.9994 & 4.564 & 0.71 & 0.016 & 0.0143 & 0.0193 & 0.7209 & 2.2 & 0.9999 \\
     	         & 5858 & 1.055 & 0.9997 & 4.555 & 0.70 & 0.017 & 0.0152 & 0.0207 & 0.7180 & 2.3 & 4.9e-5 \\
     
    \hline     
   \end{tabular}
      \label{tab:map_broad}
\end{table*}

\begin{table*}
   \centering
   \caption{Same as Table\,\ref{tab:nop} but with $R_{\rm BCZ}$, realistic HRD and narrow age priors.}
    \begin{tabular}{l c c c c c c c c c c c c} % Column formatting, @{} suppresses leading/trailing space
   \hline
   \hline
     grid & $T_{\rm eff} [K]$ & $L/L_{\odot}$ & $R/R_{\odot}$ & Age & $X_0$ & $Z_0$ & $Z_{\rm s}$ & $\ZoX$ & $R_{\rm BCZ}$ & $\alpha_{\rm ml}$  & Probability \\
     \hline
     GS98 & 5829 & 1.034 & 0.9995 & 4.591 & 0.72 & 0.016 & 0.0144 & 0.0191 & 0.7192 & 2.2 & 1.0 \\
              & 5788 & 1.004 & 0.9991 & 4.580 & 0.73 & 0.015 & 0.0135 & 0.0177 & 0.7224 & 2.1 & 1.0e-18 \\
    \hline
     AGS05 & 5795 & 1.010 & 0.9994 & 4.613 & 0.71 & 0.016 & 0.0143 & 0.0193 & 0.7204 & 2.2 & 1.0 \\
     		& 5741 & 0.974 & 0.9998 & 4.538 & 0.68 & 0.022 & 0.0199 & 0.0278 & 0.7101 & 2.2 & 5.1e-70 \\
     \hline
     AGS05N & 5809 & 1.019 & 0.9994 & 4.596 & 0.71 & 0.016 & 0.0143 & 0.0192 & 0.7220 & 2.2 & 1.0 \\

     \hline        
     AGSS09 & 5802 & 1.015 & 0.9995 & 4.580 & 0.71 & 0.016 & 0.0144 & 0.0193 & 0.7192 & 2.2 & 1.0 \\
     		   & 5762 & 0.986 & 0.9991 & 4.547 & 0.72 & 0.015 & 0.0135 & 0.0179 & 0.7227 & 2.1 & 6.8e-26 \\      	           
     \hline
     AGSS09N & 5816 & 1.024 & 0.9994 & 4.564 & 0.71 & 0.016 & 0.0143 & 0.0193 & 0.7209 & 2.2 & 0.999997 \\
     	         & 5834 & 1.039 & 0.9999 & 4.607 & 0.69 & 0.019 & 0.0171 & 0.0235 & 0.7134 & 2.3 & 2.9e-6 \\
     
    \hline     
   \end{tabular}
      \label{tab:map_real_rbc}
\end{table*}

\subsubsection{Summary}

Our detailed analysis using various priors has shown:

\begin{itemize}
\item Without priors, the frequencies fit best to models with significantly underestimated luminosities and ages of about 5 Gyrs. There is no clear preference for any specific composition.
\item Models that are constrained by the solar $L$, $T_{\rm eff}$ and $R_{\rm BCZ}$ prefer the revised composition but are still too old. Except for the age, they can reproduce all known parameters, as well as the frequencies, quite well.
\item Models that are tightly constrained by our information on the solar age suffer a strong degradation in their quality of fit. 
Depending on whether $L$ and $T_{\rm eff}$ are included as tight constraints, there is a either a clear preference for the old or the revised composition. In any case, for solar-age models $T_{\rm eff}$ is overestimated while the stellar radius is slightly underestimated, producing a significantly overestimated luminosity. The model values for $R_{\rm BCZ}$ are too high and well outside the observational uncertainties.
\end{itemize}

\section{Discussion}

In the following section, we will consider the questions formulated at the beginning of the paper. 

\subsection{The ``best fit"}

Our first question was ``Which models fit the solar modes and other observables the best?". This can be answered by looking at the evidence values for all the grids we tested in the previous section. 

Considering only the p-mode frequencies, i.e., no priors, Fig.\,\ref{fig:nop_comp} shows that GS98, GS98N, and AGSS09 all perform comparably well, as they are all able to reproduce the observed solar frequencies. Taking into account some of our prior knowledge by using the broad HRD prior, we find a similar result in Fig.\,\ref{fig:hrd_comp}. This also increases the evidence of all models. For a tighter, more realistic HRD prior the evidence for AGSS09(N) is significantly higher than for the other models.

However, we have to reject these results, as the models are clearly too old. By removing the older models via uniform age priors, we first see a large drop in the evidence values for the AGS05(N) and AGSS09(N) grids. Indeed, when we employed the narrow age prior, which is still comparatively broad (100 Myrs) to allow for systematic errors in the model evolution, the GS98(N) grid evidences suffer the same effect. We are forced to conclude that the model frequencies are getting worse as we approach the (presumably) correct solar age. A similar conclusion was reached in Paper I, but here we have shown that this is not affected by the contested different chemical compositions or the two different nuclear reaction rates.{The different compositions only produce clearly different results when using additional constraints, as discussed in the next section.

In Fig.\,\ref{fig:soundspeed} we have plotted the relative difference between the solar sound speed profile as measured from inversion \citep{basu2009} and as determined from some of our best fit models. The Model S from \cite{jcd1996} is plotted as well. This reflects our summary from above, concluding that models constrained to the 
solar observables are worse at reproducing the observed frequencies and therefore the solar sound speed profile. The figure also shows, contrary to what is commonly reported in the literature, that when using all our prior information the best GS98 model performs worse than the the models with the revised composition. It would be interesting to include the sound speed profile information in the fitting procedure as well, but the systematic differences between observations and calculations are substantial and the analysis would be non-trivial. It is therefore beyond the scope of this paper and should be targeted for future work.

\begin{figure*}
\centering
\includegraphics[width=0.75\textwidth]{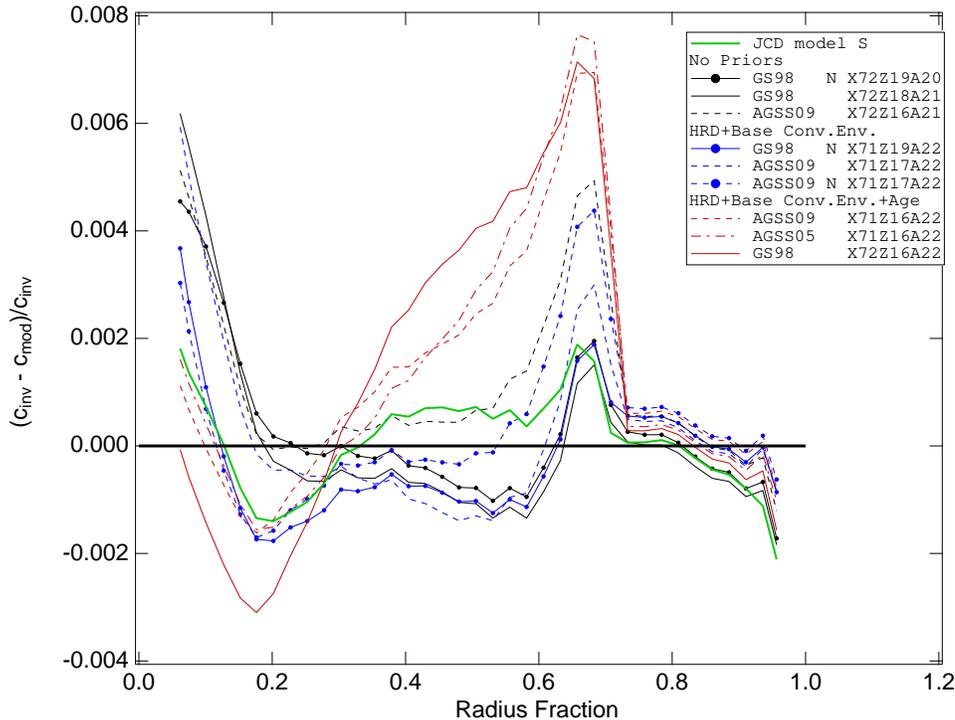}
\caption{Relative difference between the solar sound speed as measured from inversion and determined from our various models. The legend indicates which models and priors were used. Only the realistic HRD and narrow age prior results are plotted. N denotes the NACRE reaction rates. (A colour version of this figure is available in the electronic version of the paper)}
\label{fig:soundspeed}
\end{figure*}

To summarize, the argument for or against the grids presented in this paper cannot be made by simply claiming that one grid produces better frequencies in one particular setup of priors and observables. As we have shown, the grids are able to deliver similar fits in various conditions, and all grids actually have problems to fit both seismic and solar parameters. Therefore, we cannot identify a clear ``best fitting model".

\subsection{Composition}

Contrary to most studies in the literature, our results lead us to reject any clear preference for any of the contested chemical compositions over the others. Looking at the frequencies alone, no composition is clearly preferred, but there is very strong evidence against AGS05. Ignoring the model ages but using our other priors leads to a significant preference for AGSS09(N). Employing the $R_{\rm BCZ}$, narrow age, and broad HRD prior leads to decisive evidence for GS98. On the other hand, using all priors leads to a clear preference for AGS05 and AGSS09. Hence, which composition better represents the Sun depends on the consideration of tight constraints on $R_{\rm BCZ}$, $T_{\rm eff}$, $L$, and age. This suggests that our models are not calibrated well enough to the Sun, so that prior information is playing an important role compared to the observed frequencies. The latter do have an effect, however, in selecting the models that are compatible with our prior information, and thus the results cannot simply be dismissed.

We have to conclude that without solving the general problem of how to produce solar-age models that look like the Sun and produce adequate frequencies, any discussion of the contending compositions has to remain unresolved. 
Therefore, we also have to refute the claim that the AGSS09 composition is incompatible with helioseismic results.

We want to exemplify this, and contrast it to arguments used in the past, by looking at some of the solar parameters obtained from the fits. As shown in Table\,\ref{tab:map_real_rbc}, the best GS98 model when subject to all our prior knowledge constraints, has $\XO = 0.72$ and $\ZO = 0.016$, and therefore $\YO = 0.264$. For the helium mass fraction in the envelope we obtain $\Ys = 0.234$, which clearly does not agree with helioseismic inferences ($\Ys = 0.2485 \pm 0.0035$, \cite{basu2004}). Furthermore, this GS98 model over-estimates the luminosity and effective temperature. Also the location of the base of the convection zone remains a problem. For the corresponding AGSS09 model, our best model when using all prior constraints, we obtain $\XO = 0.71$ and $\ZO = 0.016$, which leads to $\YO = 0.274$. In the envelope, this then amounts to $\Ys = 0.243$, which lies almost within the $1\sigma$ uncertainties. Judging from the goodness of fit to the frequencies, as well as from the agreement to these helioseismically determined values, we would have to conclude that AGSS09 outperforms GS98. Naturally, this is only true if we ignore the overestimated luminosities, and $R_{\rm BCZ}$ values. Also, as shown in Fig.\,\ref{fig:soundspeed} both models produce some of the strongest deviations from the solar sound speed profile.

In a similar case, when ignoring the age and using just the $R_{\rm BCZ}$ and realistic HRD prior we find $\Ys = 0.2413$ and $R_{\rm BCZ} = 0.7096$ (GS98), or alternatively $\Ys = 0.2413$  and $R_{\rm BCZ} = 0.7132$ (AGSS09). 
Both models now fit the helioseismic inferences quite well, but again AGSS09 is better at reproducing the overall solar parameters and also the base of the convection zone. In this case, however, we have to consider the big problem that the ages are wrong by almost ten percent.

Consequently, we reiterate that these arguments, as well as potentially serendipitous matches of specific model properties are insufficient to solve the composition problem.

\subsection{Surface effects}
\label{sec:surf}

As mentioned in the introduction, previous studies mostly looked at frequency differences and spacings, due to the surface effect problem. We have employed the Bayesian formalism to take the surface effects into account while still using the full information provided by each frequency. 
We now want to analyse how they affect the analysis and to what extent they influence our conclusions.\footnote{Note that the surface effects are always measured with respect to specific models and using, e.g., adiabatic rather than non-adiabatic frequencies will result in different surface effects.} This is possible because, as discussed in Paper I, our method provides the most probable systematic deviations between observed and theoretical frequencies, as well as their uncertainties, for every observed mode.

One explanation for the higher evidences at older ages could be that the older models show smaller surface effects. Such a situation would pose a difficult problem, because we cannot assume that our models are already good enough in the outer layers. The determination of the basic solar parameters would again depend on the surface layers which is not what we want. Fig.\,\ref{fig:surfeffects} shows that, fortunately, the opposite is the case. It compares the surface effects as determined from our GS98 models with the ``no prior" approach to respective values from the narrow  age prior approach. The mean most probable deviation for the latter amounts to $\left<\gamma \Delta_{\rm i}\right> = -2.495$, while the narrow age prior yields $\left<\gamma \Delta_{\rm i}\right> = -2.301$. Obviously, the surface effects are larger for the ``no prior" mode, yet this model has a much higher evidence. This shows that in the case of large surface effects, the probabilities are, as required, more sensitive to the lower order modes. Fig.\,\ref{fig:surfeffects_zoom} shows these modes in more detail. In addition, we have plotted linear fits to these deviations to underline that for {\it l}=0 modes in particular, the ``no prior" approach is more consistent with the $\gamma \Delta = 0$ baseline. 

\begin{figure}
\centering
\includegraphics[width=0.8\columnwidth]{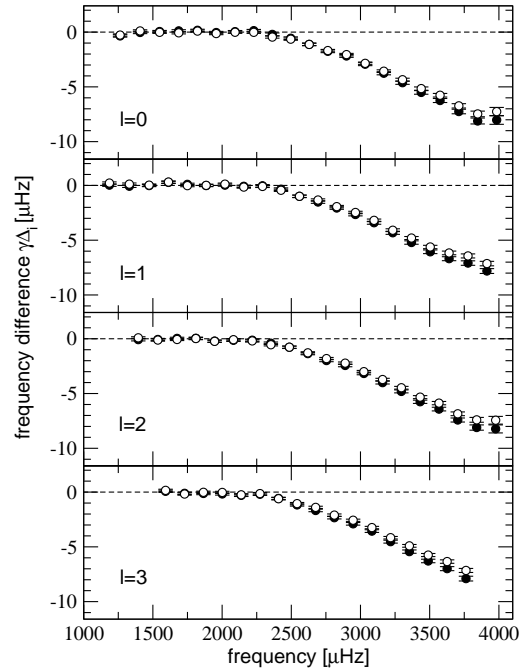}
\caption{Most probable systematic deviations and uncertainties as measured using the GS98 grid. The black circles represent the ``no prior" approach, while open circles derive from the broad HRD + narrow age prior. Note that the uncertainties are dominated by the theoretical frequency uncertainties ($0.1 \,\mu\rm Hz$) except at the highest orders. The ``no prior" approach results in larger surface effects. The dashed guide line shows a frequency difference of zero.}
\label{fig:surfeffects}
\end{figure}

\begin{figure}
\centering
\includegraphics[width=0.8\columnwidth]{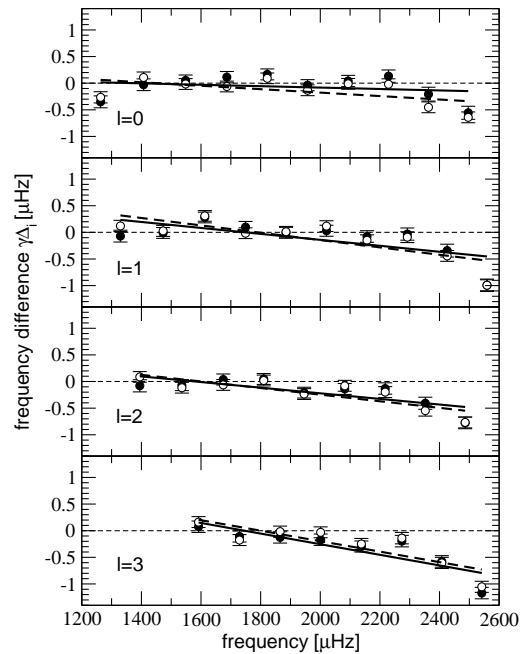}
\caption{Same as Fig.\,\ref{fig:surfeffects} but zoomed in on the lower-order modes. In addition, long dashed and solid lines represent linear fits to the open and black circles. Note that the increase in slope towards higher spherical degree is an artifact due to missing lower-order modes.}
\label{fig:surfeffects_zoom}
\end{figure} 

In a similar comparison, it is also interesting to probe the differences between the surface effects for the different compositions. Fig.\,\ref{fig:surfeffects_gsasp} and \ref{fig:surfeffects_gsasp_zoom} show a comparison of the AGSS09 and GS98 results obtained with the $R_{\rm BCZ}$, realistic HRD, and narrow age prior. The AGSS09 model exhibits a larger surface effect at the lowest orders and therefore gets penalized in the probability terms for these modes. However, it also fits better on average at the lowest $l=0$, $l=1$, and $l=3$ modes and, at the highest orders, has slightly smaller surface effects.

Comparing the systematic differences plotted in Fig.\,\ref{fig:surfeffects_zoom} and Fig.\,\ref{fig:surfeffects_gsasp_zoom} reveals that the most important component of the frequency fit is indeed the overall goodness of fit at the lower order modes. While the ``no prior" frequencies do have significantly larger surface effects above $\sim 3000\,\muHz$, the systematic differences are significantly smaller between $2000$ and $3000\,\muHz$.

In conclusion, our surface effect treatment performs favourably by allowing low-order modes to dominate the fitting process, while still being flexible enough to allow us to properly measure the most probably surface effect at higher orders for every frequency. 

\begin{figure}
\centering
\includegraphics[width=0.8\columnwidth]{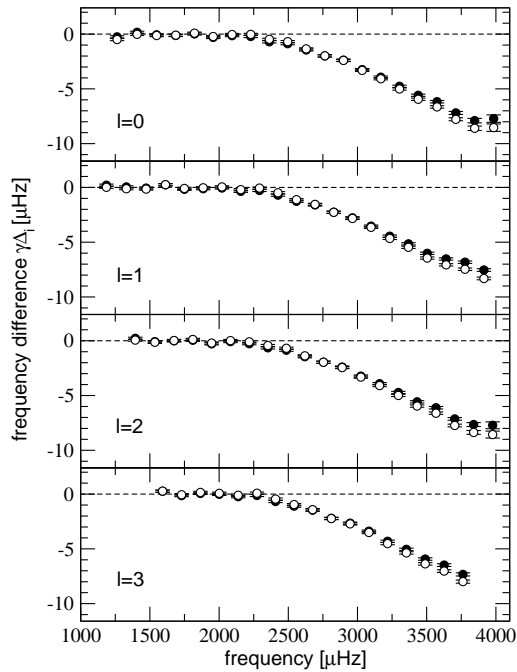}
\caption{Same as Fig.\,\ref{fig:surfeffects} but for GS98 (open circles) and AGSS09 (black circles). Both results are based on the realistic HRD + narrow age + $R_{\rm BCz}$ prior analysis.}
\label{fig:surfeffects_gsasp}
\end{figure} 

\begin{figure}
\centering
\includegraphics[width=0.8\columnwidth]{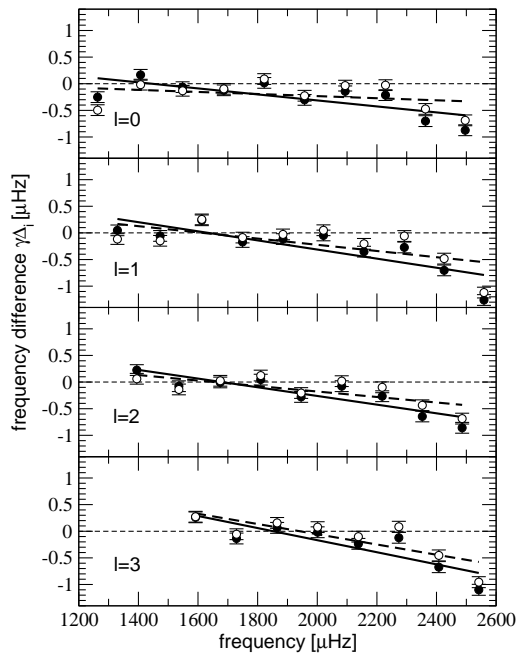}
\caption{Same as Fig.\,\ref{fig:surfeffects_gsasp} but zoomed in on the lower-order modes. In addition, long dashed and solid lines represent linear fits to the open and black circles.}
\label{fig:surfeffects_gsasp_zoom}
\end{figure} 

\subsection{Implications for asteroseismology}

General properties of Sun-like stars can now be inferred via scaling laws using high-quality frequencies from space missions and other asteroseismic observables \citep[see, e.g.,][]{huber2012}. 
While the uncertainties of the current solar frequency sets are still smaller than those of the best Kepler targets, the asteroseismology community expects to be able to go beyond scaling laws and probe details of the stellar physics (e.g., determining ages and chemical compositions). Our results suggest that in order to obtain accurate results, more work is needed to first understand the properties of the Sun. As we know from meteoritic data, we obtain solar ages that are wrong by hundreds of millions of years unless we restrict the model space. Furthermore, when we perform a full grid-based analysis, we cannot yet properly distinguish between the competing chemical compositions which have an effect on all the involved quantities. 

The impact of our analysis also extends beyond the purely asteroseismic applications. For instance, for our best models presented in this paper we obtain values for $\ZoXsun$ ranging from 0.0190 to 0.0230. If we constrain ourselves to our models at the approximate solar age, we require $\ZoXsun < 0.0205$, which is quite different from the standard value that is often used to transform between $\left[{\rm Fe/H}\right]$ and $\Zs$. In addition, uncertainties and systematic errors in the metallicity and helium abundance will naturally propagate into the results of other fields (e.g., the study of Galactic abundances) that rely on the solar calibration.

\section{Conclusions}

In this paper we have reported on our extensive grid-based ``asteroseismic" investigation of the Sun using the Bayesian formalism developed in Paper I. We extended our previous study by using different grids with competing chemical compositions (GS98, AGS05 and AGSS09) and nuclear reaction rates. We found that we cannot accurately reproduce the solar properties by fitting the frequencies alone without using prior information. On the other hand, when using prior information, we observe a strong degradation in the goodness-of-fit for the frequencies. This leads us to conclude that we cannot yet give preferential weight to either of the competing chemical compositions (or nuclear reaction rates for that matter) since the evidence values contradict our prior information. In other words, the grids are not properly calibrated and some parts of the fundamental model physics are inappropriate. Our work does not suggest that the revised compositions are any more incompatible with helioseismology in some systematic way than the traditional GS98 abundances. We have also established that it is not the outer layers which cause the problem, as our Bayesian treatment of surface effects all but removes their impact. 

The meteoritic age of the solar system of about 4.568 Gyrs is very well established (even if we allow for a systematic error of perhaps a few Myrs) and its relation to the solar model age, although not precisely known, cannot be expected to introduce a larger uncertainty than the dynamical time scales associated with evolution down the Hayashi track. Yet, if we do not constrain the solar age, we obtain values around 4.9 to 5 Gyrs, which is an error of about 10 percent. Systematic errors in the models are well below 100 Myrs, and therefore below the discrepancy between the asteroseismic solar age and the meteoritic age\footnote{The age discrepancy predates the present work. The standard, often used, reference solar model Model S by \cite{jcd1996} uses an age of 4.6 Gyrs measured from the ZAMS or approximately 4.64 Gyrs measured from the birthline.}. So although ultimately this may not be the best way to untangle our results and characterise, in a simple way, what is wrong with the models, we have come to see the problem as one related to, or at least indicated by, the age. Unfortunately, nearly every model assumption, e.g., opacities (especially of the metals), primordial abundances (especially of helium, neon, carbon, oxygen, and nitrogen), convective transport theory and the modelling of the surface convective envelope, diffusion of helium and heavier elements, winds, mass loss, magnetic fields, rotational shear at the base of the convection zone, can affect the model age.

Our conviction is that the problems reported in this paper are not caused by inadequate frequencies or the general inability to use asteroseismology in the way we have presented. Rather, we think that all the tools and the data are now at an adequate level to show us the limitations of our models. Indeed, we would like to emphasise that evidence-based Bayesian studies are an excellent way to accurately assess future developments in solar modelling. They provide a fully consistent framework to test observables, treat systematic errors (e.g., surface effects) and use prior information, in order to iterate towards more accurate model physics. Such is necessary to both better understand our Sun and to reap the full benefits of asteroseismology. 

\section*{Acknowledgments}

We would like to thank the referee for improving the quality and truly expanding the scope of this paper.
The authors acknowledge funding from the Natural Sciences and Engineering Research Council of Canada. Computational facilities were provided by ACEnet, which is funded by the Canada Foundation for Innovation (CFI),  Atlantic Canada Opportunities Agency (ACOA), and the provinces of Newfoundland and Labrador, Nova Scotia, and New Brunswick.
 
\bibliography{sunpaper}

\end{document}